\documentclass[epj,ctagsplt]{svjour}
\usepackage{epsf}
\usepackage{amsmath}
\usepackage{amssymb}

\begin{document}

\title{Properties of the asymptotic $nA+mB \to C$ reaction-diffusion
fronts.}
\author {J. Magnin}
\institute{D\'epartement de Physique Th\'eorique, Universit\'e de Gen\`eve, CH-1211 Gen\`eve 4, Switzerland.}

\date{\today}

\abstract{
We discuss, at the mean-field level, the asymptotic shape of the
reaction fronts in the general $nA+mB \to C$ reaction-diffusion
processes with initially separated reactants, thus generalizing
to arbitrary reaction-order kinetics the work done by G\'alfi and
R\'acz for the case $n=m=1$. Consequences for the formation of
Liesegang patterns are discussed.
}

\PACS{
{82.20.Mj}{Nonequilibrium kinetics} \and
{82.20.Db}{Statistical theories} \and
{66.30.Ny}{Chemical interdiffusion}
}

\maketitle


\section{Introduction}

Consider two reactants, initially separated, which are put in
contact at time $t=0$ and start to mix one into each other by
diffusion. A region where the reaction rate is high will develop
at their interface. The mathematical function describing the
variation, in space and time, of the amplitude of the reaction
rate in this region is usually called a {\it reaction front}
$R({\bf x},t)$.

Dynamical properties of reaction fronts in (purely or
effectively) one-dimensional reaction-diffusion systems
but that reduce to an effectively 
of a symmetry in the initial 
of numerous
studies~\cite{GalfiRacz,ChopCorDroz,Cornell1,Cornell2,Cornell3,Cornell4,Koza1,Koza2,Larralde,usSemiperm,Bazant1,Bazant2}. In general, it
is observed that these fronts obey {\it asymptotic scaling},
characterized by a scaling function $G$ and scaling exponents
$\alpha,\gamma$~:
\begin{equation}
  R \sim t^{-\gamma}\Phi\left(\frac{x-x_f(t)}{t^{\alpha}}\right) \quad .
\label{generRF}
\end{equation}
In the previous expression, $x_f(t)$ locates the position of the
front (usually defined as its first moment), which generally obeys
\begin{equation}
  x_f(t) \propto \sqrt{t} \quad ,
\end{equation}
accounting for the diffusive origin of the front's dynamics.

In the framework of a mean-field approximation (which is ours
from now on), scaling hypothesis, together with balance
considerations, can lead quite directly to the values of the
dynamical scaling exponents (see~\cite{GalfiRacz,ChopCorDroz}).
Accessing the structure (i.e. the {\em shape}) of the scaling
function itself requires however to go one level deeper into the
analysis of the process in consideration.

In addition to satisfying a purely theoretical curiosity, knowing
better $\Phi$ itself can provide a practical advantage. For example,
in studying phenomena involving reaction-diffusion processes,
like Liesegang patterns formation, it may be useful to dispose of
an explicit, approximated analytical form for $\Phi$.
This allows, for example, to bypass the dynamical generation of
this front in a numerical simulation and save computation time
~\cite{usmodelB}. Finding such an analytical approximation
requires evidently to gain sufficient information about the
scaling function.

Both tasks (derivation of the scaling exponents {\em and} of the
scaling function) were accomplished in the pioneering paper by
G\'alfi and R\'acz~\cite{GalfiRacz}, where they studied the
reaction front in the $A+B \stackrel{k}{\to} C$ process
with initially segregated $A$-s and $B$-s in mean-field approximation.

In the present paper, we provide a generalization of their work
to the case of arbitrary reaction-order kinetics, $nA+mB
\stackrel{k}{\to} C$, and explain how to calculate the associated
$C$ density profile in the asymptotic regime.

An important motivation for this generalization is the following~:
in the case of Liesegang patterns, the primary chemical reaction
leading (through several complex coarsening processes) to the
formation of precipitate turns out to be most often of the types
$A+2B$ or $2A+B$, and not $A+B$, as usually considered for
simplicity in theoretical models ~\cite{usMP,usmodelB,uswidth}.

\section{Scaling analysis.}

\subsection{Definition and notations.}

The case of mean-field, general reaction-order kinetics in the
initially segregated reactants case has already been addressed in
~\cite{ChopCorDroz}. Using scaling analysis, the authors showed
that the exponents controlling the asymptotic behaviour of the
reaction front are given, in terms of the reaction-order
constants, by~:
\begin{equation}
  \alpha(n,m) = \frac{n+m-1}{2(n+m+1)} \; , \;
  \gamma(n,m) = \frac{1}{n+m+1}
  \label{propscalexps}
\end{equation}
where $\alpha,\gamma$ are the same as in (\ref{generRF}), and
$(n,m)$ are integers both $\geq 1$.

It is important to note the following properties of these
exponents~:
\begin{enumerate}

\item $\alpha(n,m)=\alpha(n+m)$, \quad $\gamma(n,m)=\gamma(n+m)$.
\item $\alpha(n,m)<1/2 \; \forall \, (n,m)$,
      \quad $\alpha$ increases monotonically
      from $1/6$ $(n=m=1)$ to $1/2$ $(n+m \to \infty)$.
\item $\alpha(n,m)+\gamma(n,m)=1/2$.
\end{enumerate}

We can start from the above results to formulate a general
derivation that will lead us to the family of ordinary differential
equations defining the asymptotic shape of the reaction fronts
$R_{(n,m)}$~\footnote{This derivation follows closely the steps
and notation in~\cite{GalfiRacz}, and the reader should refer to
it for further details and justifications.}. To this goal, let us
consider the following one dimensional initial-value problem, describing a
reaction-diffusion process between initially
separated $A$ and $B$ particles in the mean-field approximation~:
\begin{subequations}
\label{RDequ-dim}
\begin{eqnarray}
\partial_T A(X,T) & = & D_A \partial_{X}^2 A(X,T) -kn(A^nB^m)(X,T) \quad , \\
\partial_T B(X,T) & = & D_B \partial_{X}^2 B(X,T) -km(A^nB^m)(X,T) \quad , \\
\partial_T C(X,T) & = & kA^n(X,T)B^m(X,T),
\end{eqnarray}
\end{subequations}
with
\begin{subequations}
\begin{eqnarray}
A(X,T=0) & = & a_0 \theta(-X) \quad , \\
B(X,T=0) & = & b_0 \theta(X) \quad , \\
C(X,T=0) & \equiv & 0.
\end{eqnarray}
\end{subequations}

\begin{sloppypar}
In the above equations,
\begin{itemize}
\item $\theta$ denotes the Heaviside step function
[$\,\theta(X<0)=0, \quad \theta(X\geq 0)=1$\,].
\item $A$, $B$ and $C$ are concentrations with dimensions
      $[A,B,C]=X^{-1}$.
\item $D_A$ and $D_B$ are diffusion coefficients
      ($[D_A,D_B] = X^{2}T^{-1}$).
\item $k$ is the reaction rate ($[k] = [X^{n+m-1}T^{-1}]$).
\end{itemize}
\end{sloppypar}

In the following, we will only consider the case of equal
diffusion coefficients, $D_A=D_B \equiv D$, since the method we
are going to use requires this strong condition to be satisfied.
The asymmetric case $D_A \neq D_B$ reveals to be several orders
of magnitude higher in difficulty. Interesting results have been
obtained in the case $n=m=1$, in connection to the front's
dynamics~\cite{Koza1,Koza2}, but a derivation of the shape of the
scaling functions for arbitrary $D_A/D_B$ seems to be still out of
reach for the moment. \\

The first step in our calculation is to render a-di\-men\-sional the
problem we are dealing with. This can be done through the
following change of variables~:
\begin{subequations}
\label{units}
\begin{eqnarray}
\label{units-X}
x & \equiv & \sqrt{\frac{ka_0^{n+m-1}}{D}} \, X \quad , \\
t & \equiv & ka_0^{n+m-1} \, T \quad , \\
a, b, c & \equiv & A/a_0, B/b_0, C/c_0 \quad .
\end{eqnarray}
\end{subequations}

The equations then read~:
\begin{subequations}
\label{RDequ-adim}
\begin{eqnarray}
\label{RDequ-a-adim}
\partial_{t} a(x,t) & = & \partial_{x}^2a(x,t) -na^n(x,t)b^m(x,t) \quad , \\
\label{RDequ-b-adim}
\partial_{t} b(x,t) & = & \partial_{x}^2b(x,t) -ma^n(x,t)b^m(x,t) \quad , \\
\label{RDequ-c-adim}
\partial_{t} c(x,t) & = & a^n(x,t)b^m(x,t) \quad ,
\end{eqnarray}
\end{subequations}
with
\begin{subequations}
\begin{eqnarray}
a(x,t = 0) & = & \theta(-x)  \quad , \\
b(x,t = 0) & = & \frac{b_0}{a_0}\theta(x) \quad ,  \\
c(x,t = 0) & \equiv & 0 \quad .
\end{eqnarray}
\end{subequations}


\subsection{Solution for $a-(n/m)b$.}

We define
\begin{equation}
u(x,t) \equiv \left(a-{n \over m}b\right)(x,t).
\end{equation}
This function obeys the diffusion equation :
\begin{subequations}
\begin{eqnarray}
\partial_t u(x,t)  &=&  \partial_x^2 u(x,t) \quad , \\
u(x<0,t=0)  &=&  1 \quad , \\
u(x>0,t=0)  &=&  -{n \over m}\frac{b_0}{a_0}
                 \equiv -{n \over m}q \quad ,
\end{eqnarray}
\end{subequations}
whose solution reads
\begin{equation}
u(x,t) = \frac{1}{2}\left( (1-{n \over m}q)-(1+{n \over m}q){\rm erf}\left(\frac{x}{2\sqrt{t}}\right)\right)  \quad .
\label{sol-u}
\end{equation}
In the above equation, ${\rm erf}$ denotes the error function, ${\rm erf}(x)
\equiv (2/\sqrt{\pi})\int_0^{x}{\exp(-w^2)dw}$.

Let $x_f(t)$ be such that $u(x_f(t),t) = 0$.
One can check that
\begin{equation}
x_f(t) = \sqrt{2D_ft} \quad ,
\end{equation}
with $D_f=D_f(q)$ given by
\begin{equation}
{\rm erf}\left(\sqrt{\frac{D_f}{2}}\right) = \frac{1-{n \over
m}q}{1+{n \over m}q} \quad .
\label{Df}
\end{equation}


\subsection{Equation for $a$ in the reaction zone.}

We write now $b = {m \over n}(a-u)$ and plug it into
(\ref{RDequ-a-adim}), getting thereby an equation for $a$
involving only $a$ and the known function $u$~:
\begin{equation}
\partial_t a(x,t)  =  \partial_x^2 a(x,t) -
                      n\left({m \over n}\right)^m
                      \Big{[}a^n (a-u)^m\Big{]}(x,t) \quad .
\label{RDequ-a-u}
\end{equation}

We are interested in the solution of this equation in the
reactive region $|x-x_f| \approx t^{\alpha(n,m)}$. As the latter
is believed to widen with a time exponent $\alpha(n,m)<1/2$ ,
this allows us to expand $u$ around $x_f$ to the lowest-order in
$x/\sqrt{t}$, since the neglected terms will vanish as $t \to
\infty$~:
\begin{equation}
u(x,t) \approx -K \frac{x-x_f}{\sqrt{t}} \quad \quad |x-x_f| \approx t^{\alpha(n,m)} \quad ,
\label{u-rz}
\end{equation}
with $K$ given by
\begin{equation}
K = \frac{1+{n \over m}q}{2\sqrt{\pi}} \exp(-D_f/2) \quad .
\label{K}
\end{equation}

The boundary conditions that the solution to (\ref{RDequ-a-u})
must satisfy in the reactive region are~:
\begin{subequations}
\begin{eqnarray}
\label{bc-a-1}
a(x \to -\infty,t) & = & -K \frac{x-x_f}{\sqrt{t}} \quad , \\
\label{bc-a-2}
a(x \to +\infty,t) & = & 0 \quad .
\end{eqnarray}
\end{subequations}


\subsection{Scaling hypothesis.}

We shall now assume that asymptotically (i.e. when $t \to
\infty$), the solution to (\ref{RDequ-a-u}) adopts the following
scaling form~:
\begin{equation}
a(x,t) \approx t^{-\gamma(n,m)} G_{(n,m)}\left(\frac{x-x_f(t)}{t^{\alpha(n,m)}}\right) \quad ,
\label{scalform-a}
\end{equation}
where $\{G_{(n,m)}\}_{n,m \geq 1}$ are a family of scaling functions remaining to be
characterized. The scaling exponents are given by
(\ref{propscalexps}).


\subsection{Differential equation for $G_{(n,m)}$.}

Let's define first the reaction zone coordinate $z$
\begin{equation}
z \equiv \frac{x-x_f}{t^{\alpha{(n,m)}}} \quad .
\end{equation}
Inside the reaction zone, $u$ and $b={m \over n}(a-u)$ write
\begin{equation}
u(z) = -Kt^{\alpha{(n,m)}-1/2}z \quad ,
\label{u(z)}
\end{equation}
\begin{equation}
b(z) =  {m \over n}(t^{-\gamma{(n,m)}}[G_{(n,m)}(z)+Kz]) \quad .
\label{b(z)}
\end{equation}

Using (\ref{scalform-a}), eq. (\ref{RDequ-a-u}) becomes~:
\begin{equation}
\begin{split}
&t^{2\alpha{(n,m)}-1}[-\gamma{(n,m)} G_{(n,m)}-\alpha{(n,m)} z \partial_z G_{(n,m)}] \\
&- \sqrt{\frac{D_f}{2}}t^{\alpha{(n,m)}-1/2} \partial_z G_{(n,m)} \\
&= \partial_z^2 G_{(n,m)} - n\left({m \over n}\right)^mG^{n}_{(n,m)}[G_{(n,m)}+Kz]^{m} \quad .
\end{split}
\end{equation}

We now take the asymptotic limit inside the reaction zone, i.e.
we let $t \to \infty$, keeping $z$ fixed. The two terms on the
left-hand side vanish (remember that $\alpha_{(n,m)}<1/2$~!) and
we remain with the following ordinary, non-linear second-order
differential equation for the scaling functions $G_{(n,m)}$~:
\begin{equation}
G_{(n,m)}''(z) =  n\left({m \over n}\right)^m G^{n}_{(n,m)}(z)[G_{(n,m)}(z)+Kz]^{m} \quad .
\label{equ-G}
\end{equation}

The boundary conditions (\ref{bc-a-1}--\ref{bc-a-2}) imply the
following asymptotics for $G_{(n,m)}$~\cite{GalfiRacz}~:
\begin{subequations}
\label{bc-G}
\begin{eqnarray}
G_{(n,m)}(z) & \to & -Kz  ,\quad z \to -\infty \quad ,  \\
G_{(n,m)}(z) & \to & 0    ,\quad z \to \infty \quad .
\end{eqnarray}
\end{subequations}

We are now left with a boundary-value problem
(\ref{equ-G},\ref{bc-G}) that can be solved numerically.


\subsection{Solving the equation for $G_{(n,m)}$.}

We can make the problem K-independent by rescaling $G$ and $z$~:
\begin{subequations}
\begin{eqnarray}
G & \equiv & K^{\mu(n,m)}\tilde{G} \quad ,
\\ z & \equiv & K^{\nu(n,m)}\tilde{z} \quad ,
\end{eqnarray}
\end{subequations}
and by using a suitable choice for $\mu$ and $\nu$. Inserting
these scaled forms into (\ref{equ-G}) and imposing that K drops
out leads to
\begin{subequations}
\begin{eqnarray}
\mu(n,m) & = & \frac{2}{n+m+1} = \mu(n+m) \quad , \\
\nu(n,m) & = & \frac{-(n+m-1)}{n+m+1} = \nu(n+m) \quad .
\end{eqnarray}
\end{subequations}

The problem we are left to treat is now~:
\begin{subequations}
\label{Gresc-problem}
\begin{eqnarray}
\label{Gresc-problem-1}
\tilde{G}''_{n,m}(\tilde{z}) & = & n\left({m \over n}\right)^m
                                   \tilde{G}_{n,m}^n(\tilde{z})
                                   \Big{[}\tilde{G}_{n,m}(\tilde{z})+\tilde{z}\Big{]}^m \quad ,
                                   \\
\label{Gresc-problem-2}
\tilde{G}_{n,m}(\tilde{z}) & \to & -\tilde{z} \quad\quad \tilde{z} \to
\infty  \quad , \\
\label{Gresc-problem-3}
\tilde{G}_{n,m}(\tilde{z}) & \to & 0 \quad\quad
\tilde{z} \to \infty  \quad .
\end{eqnarray}
\end{subequations}

The reader should keep in mind, from now on, that the ``tilde''
sign stands for quantities expressed in terms of the rescaled,
$K-$independent version of the scaling function $G_{(n,m)}$ and
reactive coordinate $z$.

\subsection{The dimensionless reaction front.}

By definition~:
\begin{eqnarray}
R_{(n,m)}(x,t) & = & a^n(x,t)b^m(x,t) \nonumber \\
               & = & \left({m \over n}\right)^m t^{-\frac{n+m}{n+m+1}}G_{(n,m)}^n[G_{(n,m)}+Kz]^m \nonumber \\
               & \equiv & t^{-\beta(n,m)} F_{(n,m)}(z)  \quad ,
\label{def-R}
\end{eqnarray}
the last inequality defining both the reaction rate amplitude exponent $\beta$
and the asymptotic reaction front scaling function $F_{(n,m)}$.

\begin{figure}[htb]
\centerline{
        \epsfxsize=8cm
        \epsfbox{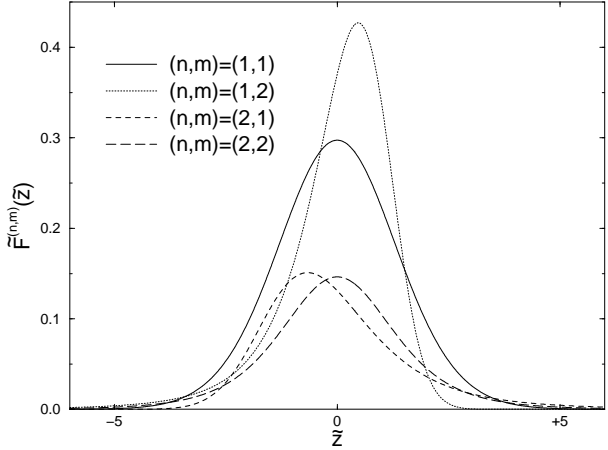}
           }
\caption{Numerical solutions to
(\ref{Gresc-problem}) for $(1 \leq n,m \leq 2)$.}
\label{rfronts}
\end{figure}

Figure \ref{rfronts} shows the result of the numerical computation
of the fronts $\tilde{F}_{(n,m)} \equiv F{(\tilde{G},\tilde{z})}$
for the cases $2\leq n+m\leq 4$ (only such low values of $n+m$ are
relevant in connection to experiments). The reader
should not be surprised by the asymmetry between the
$(n,m)=(1,2)$ and $(n,m)=(2,1)$ cases~: it is due to the passage
from the variables $\{G_{(n,m)},z\}$ to
$\{\tilde{G}_{(n,m)},\tilde{z}\}$, since both quantities are
rescaled by a $K-$ dependent factor which is {\em not}
$(n,m)$-symmetric ! \\

These fronts are all essentially localized around $z=0$, as Fig.
\ref{rfronts} suggests. In fact, one can check that in the
simplest $(n,m)=(1,1)$ case, the dominant contribution to both
tails at $z \to \pm \infty$ is proportional to $\exp (-z^{3/2})$.
In the cases $n=1,m>1$, the decay can be shown to be dominated by
$\exp (-z^{1+m/2})$ at $z \to \infty$ but only algebraically for
$z \to -\infty$ : $R(z \to -\infty) \approx z^{-(2m+1)/(m-1)}$.
Finally, when $n>1,m>1$, both tails are algebraic, with $R(z \to
\infty) \approx z^{-(2m+n)/(n-1)}$.

It is also worth noting that in the asymmetric cases $n \neq m$,
$x_f$ does not coincide with the maximum of the front.


\section{The $C$ concentration profile.}
\label{calcc0}

\subsection{Derivation of the asymptotic profile.}

We are interested now in estimating the (possibly x-de\-pen\-dent)
density $c_0^{(n,m)}(x)$ of $C$ particles left behind by the
fronts $R_{(n,m)}$ which travel diffusively through the system.
This quantity is, for example, of great importance in the theories of
Liesegang pattern formation~\cite{usMP,usmodelB}.

In dimensionless units, $c_0^{(n,m)}$ is formally given by
\begin{equation}
c_0^{(n,m)}(x) \equiv \int\limits_0^{\infty}{R_{(n,m)}(x,t) \, dt} \quad .
\label{defc0}
\end{equation}

\begin{sloppypar}
Due to the several timescales dependence of $R_{(n,m)}$, this
integral is difficult to handle. The estimation of $c_0^{(n,m)}(x)$
turns out however to be possible by making use of
\end{sloppypar}
\begin{enumerate}
\item [a)] the precious algebraic relation $\alpha+\gamma=1/2$ between the
scaling exponents, and
\item [b)] the particular structure of the solutions to
(\ref{Gresc-problem}).
\end{enumerate}

Let's consider first a narrow slice $\delta F_{(n,m)}(z_0,\delta w
)$ of the scaling function $F_{(n,m)}$, centered on $z=z_0$, of
width $\delta w \ll 1$. In the spirit of the Riemann integral, we
can approximate the amplitude of $\delta F$ inside $[z_0-\delta
w/2,z_0+\delta w/2]$ by its value $F_{(n,m)}(z_0)$ at the center.

We can estimate the contribution $\delta C^{(n,m)}(z_0,x)/\delta
x$ of this slice to the asymptotic local $C$ density inside
$[x,x+\delta x]$ as follows. The fraction of the front we are
considering will reach $x$ at a certain time $t(z_0,x)$. The
quantity of $C$ particles deposited in the interval $[x,x+\delta
x]$ will be proportional to the amplitude, the width and inversely
proportional to the speed of the slice at $t=t(z_0,x)$~:
\begin{eqnarray}
  \lefteqn{\delta C^{(n,m)}(z_0,x) \approx} \nonumber \\ &&
  \frac{t(z_0,x)^{-\gamma{(n,m)}}F_{(n,m)}(z_0)t(z_0,x)^{\alpha{(n,m)}}\delta w}
       {\sqrt{D_f/(2t(z_0,x))}} \, \delta x \nonumber \\
       & = & \sqrt{2 / D_f} F_{(n,m)}(z_0) \delta w \, \delta x  \quad .
\end{eqnarray}

In other words, the contribution of the slice to the density at
$x$ is proportional to its "mass" $F_{(n,m)}(z_0) \delta w$ but
independent of $t(z_0,x)$, and hence of $x$. This indicates that
the asymptotic density profile is flat. By superposition, our
argument leads immediately to the result we are looking for~:
\begin{eqnarray}
  c_0^{(n,m)}(x) & \equiv & c_0^{(n,m)} = {\rm const.} \nonumber \\
                 & \approx & \sqrt{2 / D_f}  {\int_{-\infty}^{\infty}{F_{(n,m)}(z) \, dz}} \quad .
\end{eqnarray}

Now our real fortune is that we are able to evaluate
analytically $\int{F_{(n,m)}}$. Using (\ref{equ-G}) and (\ref{def-R}), We have
\begin{equation}
\begin{split}
\int_{-\infty}^{\infty}{F_{(n,m)}(z) \, dz}
    = {1 \over n} \Big[ &\partial_zG_{(n,m)}(z \to \infty) - \\
    &\partial_zG_{(n,m)}(z \to -\infty) \Big] \quad .
\end{split}
\end{equation}

It is intuitively clear, from purely physical considerations,
that the solution
will converge to its
values at $\pm \infty$ in such a way that
\begin{equation}
\lim_{z \to -\infty}G_{(n,m)}'(z) = -K \; , \;
\lim_{z \to +\infty}G_{(n,m)}'(z) = 0
\end{equation}
[see (\ref{Gresc-problem-2})--(\ref{Gresc-problem-3})]. So we finally have
\begin{equation}
\int_{-\infty}^{\infty}{F_{(n,m)}(z) \, dz} = {K \over n} \quad ,
\end{equation}
and we end up with the result
\begin{equation}
c_0^{(n,m)} \approx \frac{K}{n} \sqrt{2 / D_f} \quad .
\label{c0-adim}
\end{equation}

Going back to the dimensional variables $A,B,C,X$ and $T$, one
can check that (\ref{c0-adim}) writes~:
\begin{equation}
  c_0^{(n,m)} \approx \sqrt{\frac{2D}{D_f}}{ Ka_0 \over n} \quad .
  \label{c0-dim}
\end{equation}

\vspace{0.8cm}
\begin{figure}[htb]
\centerline{
        \epsfxsize=8cm
        \epsfbox{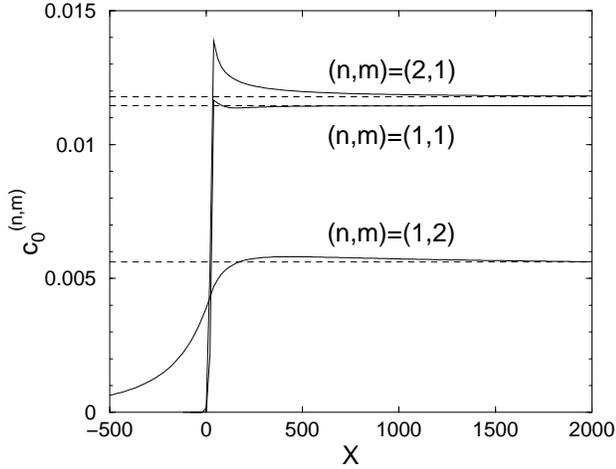}
           }
\caption{Numerical solutions to (\ref{RDequ-dim})
 for $(2 \leq n+m \leq 3)$ (plain curves). The dashed lines
 indicate the asymptotic values as calculated from
 (\ref{Df}), (\ref{K}) and (\ref{c0-dim}). The following values of the
 "free" parameters were chosen for all three cases~: $D=1$,
 $a_0=10^{-2}b_0=1$, $k=0.1$. Units for $X$ and $k$ vary with $(n,m)$
 according to
 (\ref{units-X}), while $c_0^{(n,m)}$ is measured in units of $1/a_0$.}
 \label{c0-profiles}
\end{figure}
\vspace{0.8cm}

Fig. \ref{c0-profiles} shows the $c_0^{(n,m)}$ profiles obtained
by numerical integration of the reaction-diffusion equations
(\ref{RDequ-dim}), together with the asymptotic values predicted
by (\ref{c0-dim}). \\

\subsection{Low $q$ expansion.}

In the context of Liesegang patterns-forming experiments, which
are generally based on the penetration of a highly concentrated
solution into a dissolved one, it is useful to dispose of an
expansion of this result for low $q \equiv b_0/a_0$ values. We
recall that $D_f$ is given by

\begin{equation}
{\rm Erf}(\sqrt{\frac{D_f}{2D}}) = \frac{1-{n \over m}q}{1+{n
\over m}q} = 1-2{n \over m}q \stackrel{q \ll 1, n \leq 2}{\cong} 1 \quad .
\label{erfdef}
\end{equation}

The large $x$ asymptotics of ${\rm Erf}(x)$ is given
by~\cite{GradRyzh}~:
\begin{equation}
{\rm Erf}(x) \sim 1 - \frac{e^{-x^2}}{\sqrt{\pi}x} \Big{[}
1-\frac{1}{2x^2}+ \cdots \Big{]} \quad ,
\end{equation}
so we obtain from (\ref{erfdef})~:
\begin{equation}
\begin{split}
\sqrt{\frac{D}{D_f}}\exp(-D_f/2D) \cong &\Big{[} 1- \frac{1}{2}\frac{D}{D_f} \Big{]}^{-1} \cdot \\
&\cdot \sqrt{2\pi}\frac{{n \over m}q}{1+{n \over m}q} \quad ,
\end{split}
\end{equation}
which, together with the dimensional expression for $K$,
\begin{equation}
K = \frac{1+{n \over m}q}{2\sqrt{\pi}} \exp(-D_f/2D) \quad ,
\end{equation}
gives finally
\begin{equation}
  c_0^{(n,m)} \cong
  {b_0 \over m} \left[ 1+\frac{1}{2}\frac{D}{D_f} + {\mathcal O}\left((\frac{D}{D_f})^2\right) \right] \quad \quad \nonumber \quad .
  \label{c0result}
\end{equation}

The physical meaning of this result is clear: if $D_A=D_B=D$ and
$q \ll 1$, then $D_f \gg D$ and the $B$ particles appear as
nearly immobile for the invading $A$-s. As $m$ $B$-s are required to
produce one $C$, the density equals $b_0/m$ to a very good
approximation.

However, in the typical conditions of a Liesegang experiment (
where $D_A \cong D_B$ and $10^{-2} \leq q \leq 5 \cdot 10^{-2}$
typically), the first order correction in $D/D_f$ to
$c_0^{(n,m)}$ lies in the range $(0.1-0.2)(b_0/m)$, and should
therefore, in principle, not be neglected, as can be seen of
Fig.~\ref{c0-profiles}.

\section{Summary.}

We have derived the family of ordinary differential equations
defining the asymptotic shape of the reaction fronts in the
$nA+mB \stackrel{k}{\to} C$ reaction-diffusion process with
initially separated reactants [eqs.~(\ref{equ-G}),(\ref{bc-G})].
The four lowest-order cases in $n+m$ have been solved numerically
(Fig.~\ref{rfronts}). We have also shown, and confirmed by
numerical simulations, that the density $c_0^{(n,m)}$ of $C$
particles deposited in the system by these traveling fronts is
asymptotically constant~(Fig. \ref{c0-profiles}), and we have
made explicit the dependence of this density on the reaction
orders $n,m$, as well as on the material parameters $D, a_0, k$
and $b_0$ entering the problem~[eq.~(\ref{c0-dim})].

\section{Conclusion.}

To conclude this study, we would like to comment on the
interesting phenomenon shown by Fig.~\ref{c0-profiles}~: when,
for the two $m=1$ cases, no significant quantity of reaction
product is created in the majority species subspace, the case
where $m=2$ exhibits, on contrary, an important deposit of $C$ on
the left hand side, up to far beyond the initial location of the
interface. This fact must be evidently related to the details of
the short-time dynamics of the reaction front. Some studies have
already been carried on the early-time regime subject for the
$n=m=1$ case~\cite{Taitelbaum1,Taitelbaum2} in the past. They
unveiled the existence of a surprisingly complex behaviour,
including successive power-law regimes for the early front's
dynamics, and even the possibility of a change in its direction
of motion. Such nontrivial behaviour has also been observed
numerically in the higher-order kinetics cases we have addressed
in the present paper, and a detailed study of the dependence on
$(n,m)$ of the short-time dynamics should be the object of a
forthcoming paper.


\section{Acknowledgments}

I thank Michel Droz for useful remarks.

\end{document}